\documentclass[aps,twocolumn,superscriptaddress,prl]{revtex4}
\usepackage{amsmath,latexsym,epsf}
\usepackage{graphicx}
\usepackage{subfigure}

\begin{document}

\title{Semiclassical theory for quantum spin  models with ring exchange on the triangular lattice}

\author{Amulya V Madhav}

\affiliation{Kavli Institute for Theoretical Physics, University of California at Santa Barbara, CA 93106}

\date{\today}

\begin{abstract}

A semiclassical theory of a quantum spin$-S$ model with competing ring and Heisenberg exchange terms on the triangular lattice is obtained. A mechanism for the generation of $Z_2$ vortices is exhibited. The vortices are shown to carry a nontrivial geometric phase for the order parameter when $2S$ is odd, leading to a difference between the quantum disordered ground states and low energy spectra for half odd integer and half even integer spin systems, and a topological degeneracy on surfaces with nontrivial cycles. A connection to dimer models is discussed.

\end{abstract}

\vspace{0.15cm}

\maketitle

Since the original proposal \cite{fazekspwa,pwa} that frustrated quantum spin half systems on the triangular lattice may fail to exhibit any form of local order at zero temperature, many efforts have been made to characterize such spin liquid states \cite{spinliqgenrefs}. Quantum spin (and bosonic) systems with global $XY$ symmetry have been argued to possess disordered ground states with a proliferation of $Z_2$ vortices \cite{senthilfisher}, as have dimer models \cite{sondhi,moessner} and $Sp(N)$ models for large$-N$ \cite{sachdev1}. Exact diagonalizations \cite{lhuillier,misguich} of the $SU(2)$ invariant spin half model with ring exchange interactions on the triangular lattice indicate the lack of any local order or lattice symmetry breaking in the thermodynamic limit but an understanding of the microscopic origin of the disorder in such spin models is incomplete. In this paper a semiclassical approach to the low energy physics of the ring exchange model and similar $SU(2)$ invariant systems with noncollinear order is presented \cite{sachdev1}, and the microscopic origin of $Z_2$ vortices which disorder the system is elucidated. A geometric phase associated with the vortices, arising from a topological term in the effective action for spin $S$, is argued to give rise to basic distinctions between the low energy spectra for the cases when $2S$ is odd or even.

The Hamiltonian of the ring exchange model for spin half is

\begin{eqnarray}
\label{ringham}
H &=& J_2 \sum_{<ij>} ({\bf S}_i \cdot {\bf S}_j) + J_4 \sum_{ijkl}[ 4({\bf S}_i \cdot {\bf S}_j)({\bf S}_k \cdot {\bf S}_l) \\ \nonumber
&&+4({\bf S}_i \cdot {\bf S}_l) ( {\bf S}_j \cdot {\bf S}_k)-4({\bf S}_i \cdot {\bf S}_k)({\bf S}_j \cdot {\bf S}_l) ]
\end{eqnarray}

where $<ij>$ denotes that the sites are nearest neighbours and in the second term the sum is over all elementary plaquettes bounded by sites $ijkl$ on the triangular lattice. These interactions  have been used to describe $He^{3}$ monolayers on graphite \cite{heliumlayr} and are generated in order $t^4/U^3$ in the $t/U$ expansion of the half filled Hubbard model. Since each spin is shared by twelve rhombi the spins are highly frustrated. This model will be studied for large $S$; to keep the ratio $J_4/J_2$ fixed in this limit $J_4$ must be scaled by $S^2$.

  A low energy theory for this model is obtained by treating the effects of short range fluctuations at tree level (to order in $1/S$), while maintaining the nonlinearity of the slowly varying fluctuations of the local classical order and accounting for the Berry phases which arise in the path integral for the time evolution operator \cite{haldane1, haldane1a}. It was shown in \cite{kubomomoi} that the classical limit of the ring exchange Hamiltonian (\ref{ringham}) has a four sublattice "tetrahedral" ground state on the triangular lattice for $|J_4/J_2| \le 2$. The spins on the vertices of an elementary rhombus $A_1 A_2 A_3 A_4$ are given in terms of an orthonormal triad ${\bf e}_{A}, {\bf e}_B, {\bf e}_C$ by 
\begin{eqnarray}
&&{\bf S}_{A_1}=\frac{S}{\sqrt{3}}({\bf e}_{A}+{\bf e}_{B}+{\bf e}_{C}), \\ \nonumber 
&&{\bf S}_{A_2}=\frac{S}{\sqrt{3}}({\bf e}_{A}-{\bf e}_{B}-{\bf e}_{C}), \\ \nonumber 
&&{\bf S}_{A_3}=\frac{S}{\sqrt{3}}(-{\bf e}_{A}+{\bf e}_{B}-{\bf e}_{C}), \\ \nonumber  
&&{\bf S}_{A_4}=\frac{S}{\sqrt{3}}(-{\bf e}_{A}-{\bf e}_{B}+{\bf e}_{C});
\end{eqnarray}
 the spins point to the vertices of a regular tetrahedron. This state is invariant under global $O(3) \times O(3)_{spin}$, under which the rotation matrix ${\bf e}_{\alpha,a}$, $(\alpha=A$, $B$, $C$; $a=x$, $y$, $z)$, transforms as ${\bf e}^{'}=L^{T} {\bf e} R$, with $L$, $R \in O(3)$; the additional $O(3)$ symmetry is special to the tetrahedral state; planar states in general possess  $O(2) \times O(3)_{spin}$ symmetry while an arbitrary nonplanar spiral state possesses only the spin symmetry.

 The path integral for the time evolution operator, $Z(T) \equiv \big{<} {\bf \Omega} | \exp (-i H T) | {\bf \Omega} \big{>}$, may be obtained using spin coherent states \cite{haldane1} by insertion of $N$ resolutions of the identity in the matrix element for the time evolution operator in the coherent state basis, taking the limit of infinite $N$ assuming that only continuous paths on the unit sphere contribute, and is given by

\begin{eqnarray}
Z=\int D \hat{{\bf \Omega}} e^{-\frac{i}{\hbar} \int dt H_S (t)} e^{iS \omega(\hat{{\bf \Omega}})},
\end{eqnarray}
 
where the path integral is over closed paths in time, since it is obtained from diagonal matrix elements in the spin coherent state basis, $ \hat{{\bf \Omega}_{m,n}}={\bf S}_{m,n}/S$ is the normalized spin at site ${\bf x}_{m,n}=m {\bf u}_1+n {\bf u}_2$,with ${\bf u}_{1,2}$ being elementary lattice vectors, $H_S$ is the Hamiltonian (which may be arranged as series in $S^{-1}$) and  $ \omega(\hat{{\bf \Omega}})$ is the solid angle subtended by the closed path traced out on the unit sphere by the time evolution of $\hat{{\bf \Omega}}$. The solid angle $ \omega(\hat{{\bf \Omega}})$ does not have an explicitly rotationally invariant local expression in $\hat{{\bf \Omega}}$; it is the action due to a monopole of strength $2S$ at the center of the unit sphere on which $\hat{{\bf \Omega}}$ lies. 

In the semiclassical limit the triad ${{\bf e}_{A}, {\bf e}_B, {\bf e}_C}$ is allowed to vary spatially; the spin is 
\begin{equation}
{\bf S}_{m,n}=S {\bf e}_{1,m,n}+a {\bf L}_{m,n}-\frac{a^2 {\bf L}^2 {\bf e}_{1,m,n}}{2S}+O(S^{-2}),
\end{equation}
 where we define the orthonormal vectors 
\begin{eqnarray}
{\bf e}_{1,m,n}&=&\frac{1}{\sqrt{3}} ((-1)^m {\bf e}_{A,m,n}+(-1)^n {\bf e}_{B,m,n}+\\ \nonumber
&&(-1)^{m+n} {\bf e}_{C,m,n}), \\ \nonumber
{\bf e}_{2,m,n}&=& (\frac{(-1)^{m+n}}{\sqrt{6}} {\bf e}_{A,m,n}+\frac{1}{\sqrt{6}} {\bf e}_{B,m,n} \\ \nonumber 
&&-(-1)^{m+n} \sqrt{\frac{2}{3}} {\bf e}_{C,m,n}), \\ \nonumber 
{\bf e}_{3,m,n}&=&\frac{(-1)^{m+n}}{\sqrt{2}} (-{\bf e}_{A,m,n}+{\bf e}_{B,m,n}).
\end{eqnarray}
The short range fluctuation vector ${\bf L}$ is normal to ${\bf e}_1$; thus 
\begin{eqnarray}
{\bf L}&=&((-1)^m L_1+(-1)^n L_2+(-1)^{m+n} L_3) {\bf e}_{2,m,n} \\ \nonumber 
&&+(L_{\perp,1}+(-1)^{m+n} L_{\perp,2}){\bf e}_{3,m,n},
\end{eqnarray}

 where the vector ${\bf L}$ has been separated into components which vary rapidly in space with different frequencies. The action $S_T=S \sum \omega ({\bf S}_{m,n})-\oint H_S dt$ is expanded to quadratic order in ${\bf L}$ and $\nabla {\bf e}$, higher powers of ${\bf L}$ being lower order in $S$. The Berry phase term is 
\begin{eqnarray}
S\omega({\bf S}_{m,n}/S)&=& S\omega({\bf e}_{1,m,n})+\\ \nonumber 
&&\oint dt {\bf L}_{m,n} \cdot \partial_t {\bf e}_{1,m,n} \times {\bf e}_{1,m,n}+O(1/S).
\end{eqnarray}
 Thus, the spin path integral  is

\begin{eqnarray} 
Z(T)=\int D {\bf e}_{A,B,C} D {\bf L} \exp (i S \sum_{m,n} \omega ({\bf e}_{1,m,n})+i S_d)
\end{eqnarray}

where $S_d=a \sum_{m,n} \oint dt [{\bf L}_{m,n}\cdot \partial_t {\bf e}_{1,m,n} \times {\bf e}_{1,m,n}-H_S ({\bf e}, \nabla {\bf e}, {\bf L})]$. Since ${\bf L}$ appears quadratically in the action, it can be integrated out to obtain an effective action for the fields ${\bf e}_{A,B,C}$; terms in $S_d$ which vary with momenta near $\pi$ times the reciprocal lattice vectors average to zero. In fact the $O(3) \times O(3)_{spin}$ symmetry determines the sigma model action upto constants; on the lattice it is 
\begin{eqnarray}
S_d&=&\oint dt \sum_{{\bf x}_{m,n}, {\bf \delta}, {i=ABC}} \frac{1}{2 g_0^2} [ |\partial_t {\bf e}_{i} ({\bf x}_{m,n})|^2 +\\ \nonumber
&& v_0^2 {\bf e}_i ({\bf x}_{m,n}) \cdot {\bf e}_i ({\bf x}_{m,n}+{\bf \delta})]
\end{eqnarray}
 where the (bare) couplings $g_0^2$ and $v_0^2$ are related to $J_2$, $J_4$ as above.These coupling will however flow with the length scale; in the following the effective couplings will be referred to.

The evaluation of the Berry phase $\sum_{m,n} \omega ({\bf e}_{1,m,n})$ is a special case of the result of \cite{avmthesis, avmhaldane} and is done as follows. Define the vectors ${\bf f}({\bf x}_{m,n},u)$ and ${\bf g} ({\bf x}_{m,n},v)$, which interpolate continuously on the unit sphere on the great circles between ${\bf e}_{1,m,n}$ and ${\bf e}_{3,m,n}$, and between ${\bf e}_{3,m,n}$ and the slowly varying vector ${\bf e}_{C,m,n}$, by 
\begin{eqnarray}
&&{\bf f}({\bf x}_{m,n},u)={\bf e}_{1,m,n} \sin u + {\bf e}_{3,m,n} \cos u \\ \nonumber  
&&{\bf g} ({\bf x}_{m,n},u)={\bf e}_{3,m,n} \sin u+ {\bf e}_{C,m,n} \cos u.
\end{eqnarray}
 The Berry phase at a given site, the solid angle $S \omega({\bf e}_{1,m,n})$, can be written as the solid angle swept out by the orthogonal vector ${\bf e}_{3,m,n}$, $S\omega({\bf e}_{1,m,n})$, plus the solid angle swept out by ${\bf e}_{1,m,n}$ in a frame comoving with ${\bf e}_{3,m,n}$, i.e,
\begin{eqnarray}
&&S \omega({\bf e}_{1,m,n})=S\omega({\bf e}_{3,m,n})+ \\ \nonumber 
&& S \oint dt \int_{0}^{\frac{\pi}{2}}\frac{\partial {\bf f} ({\bf x}_{m,n},u)}{\partial t} \cdot {\bf f} ({\bf x}_{m,n},u) \times \frac{\partial {\bf f} ({\bf x}_{m,n},u)}{\partial u}.
\end{eqnarray}
 Similarly, the solid angle swept out by the vector ${\bf e}_{3,m,n}$ can be expressed in terms of the solid angle swept out by the slowly varying vector ${\bf e}_{C,m,n}$ orthogonal to ${\bf e}_{3,m,n}$ using the interpolating vector ${\bf g}$ to obtain for the solid angle for ${\bf e}_{1,m,n}$ an expression in terms of the slowly varying triad ${\bf e}_{A,B,C}$; 
\begin{eqnarray}
&&S \omega({\bf e}_{1,m,n})=S\omega({\bf e}_{C,m,n})+\\ \nonumber 
&& S \oint dt \int_{0}^{\frac{\pi}{2}} [\frac{\partial {\bf g} ({\bf x}_{m,n},u)}{\partial t} \cdot {\bf g} ({\bf x}_{m,n},u) \times \frac{\partial {\bf g} ({\bf x}_{m,n},u)}{\partial u}+\\  \nonumber
&& \frac{\partial {\bf f} ({\bf x}_{m,n},u)}{\partial t} \cdot {\bf f} ({\bf x}_{m,n},u) \times \frac{\partial {\bf f} ({\bf x}_{m,n},u)}{\partial u}]
\end{eqnarray}
 and integrating over $u$, the Berry phase is 
\begin{eqnarray}
&&S\omega({\bf e}_{1,m,n})=S \omega({\bf e}_{C,m,n})-\\ \nonumber 
&&S \oint dt \frac{1}{2} \big{(} {\bf e}_{A,m,n} \cdot \frac{\partial {\bf e}_{B,m,n}}{\partial t}-{\bf e}_{B,m,n} \cdot \frac{\partial {\bf e}_{A,m,n}}{\partial t} \big{)}.
\end{eqnarray}
 As shown in \cite{avmthesis} this quantity is a topological invariant; its variation under an arbitrarily small change in the triad ${\bf e}$ vanishes. If $\Lambda \{ {\bf e}_{A,B,C} \}$ is defined by
\begin{eqnarray}
&&\Lambda \{ {\bf e}_{A,B,C} \} \equiv \exp \big{ \{ } \frac{i}{2}[ \omega({\bf e}_{C,m,n})- \\  \nonumber 
&&\oint dt \frac{1}{2}({\bf e}_{A,m,n} \cdot \frac{\partial {\bf e}_{B,m,n}}{ \partial t}-{\bf e}_{B,m,n} \cdot \frac{\partial {\bf e}_{A,m,n})}{\partial t}] \big{ \} }
\end{eqnarray} 
 then it can be seen that \cite{avmthesis} $\Lambda \{ {\bf e}_{A,B,C} \} \in \pi_1 (SO(3))$, where $\pi_1 (SO(3))$ is the fundamental group $Z_2$ of $SO(3)$. The spin path integral thus takes the form
\begin{eqnarray}
Z(T)=\int D {\bf e}_{A,B,C} \prod_{m,n} [\Lambda \{ {\bf e}_{A,B,C}({\bf x}_{m,n}) \}  ]^{2S} \exp (i S_d).
\end{eqnarray}

Consider in a given spin configuration entering in the path integral the spins in a line along a lattice vector. Labelling the time history of each spin along the line by  $\Lambda \{ {\bf e}_{A,B,C} \}$ for that site, a change of  $\Lambda \{ {\bf e}_{A,B,C} \}$ from one spin to the next requires a breakdown of continuity of the triad ${\bf e}_{A,B,C}$ in spacetime. Analogously, introducing the spatial $Z_2$ character along the chosen line at different time slices it can be seen that the breakdown of the triad occurs at a point in time. The point in spacetime where the breakdown of continuity occurs is the location of a ($1+1$ dimensional) instanton. When $2S$ is odd, the term $ \prod_{m,n} [\Lambda \{ {\bf e}_{A,B,C}({\bf x}_{m,n}) \}]^{2S}$ appearing in the path integral associates a phase $\pi$ to the rotation of the local triad ${\bf e}_{A,B,C}$ about an axis in configuration space when the triad is transported around a loop enclosing the $Z_2$ instanton. In $2+1$ dimensions, by continuing the instanton for the line in transverse directions, it is seen that $Z_2$ vortices are present and that the phase holonomy of the local triad around any loop enclosing $n_v$ vortices is $2 \pi n_v S$. Since the triads must describe closed paths in configuration space in the coherent state path integral formulation, the vortex loops must be closed in the absence of sources. 

Thus when $2S$ is even the effective theory is that of the $O(3)$ principal chiral model, which bears many similarities to the $O(3)$ nonlinear sigma model \cite{polyakov}. For sufficiently small coupling $g_0$ the principal chiral model describes long wavelength fluctuations about the tetrahedrally ordered state. In the ordered state the energy of a vortex loop is proportional to its length and to the spin stiffness. When $g_0$ exceeds a critical value $g_c$ the spin stiffness vanishes, a gap opens up for local excitations and the vortex loops proliferate; their effects on the disordered state depend also on the phase factors associated with them.

 When the system is placed on a two dimensional surface of genus $g$ (e.g. by choosing appropriate boundary conditions) in the disordered phase the ground state degeneracy is determined by the $Z_2$ character $ \Pi_{m,n} [\Lambda \{ {\bf e}_{A,B,C}({\bf x}_{m,n}) \}]^{2S}$. Order parameter configurations which have a vortex which winds along a nontrivial cycle of the surface are not continuously deformable to those which do not contain the vortex; thus the classical configurations fall into $2g$ distinct sectors. Tunnelling between sectors can occur when a vortex pierces the surface. Consider such a $Z_2$ vortex tunnelling between two such sectors through a plaquette; more complicated configurations of the vortex can be treated by decomposing the vortex into a sum of closed (contractible) loops and the above . Translation of the vortex by a lattice site in a  transverse direction to the vortex leads to a phase change $2 \pi S$ in the path integral due to the phase for the triads associated with the vortex, but no change in energy. When $2S$ is odd there is a cancellation in the path integral between these configurations differing by the translation of the vortex by a lattice vector, and thus vortices are excluded from the system. This leads to a twofold degeneracy associated with each cycle and an overall degeneracy of the ground state of $2^{2g}$ when $2S$ is odd.

If it is assumed that the disordered phase for $g > g_c$ can be continued to large $g$ without a phase transition, the action can be studied in an expansion in $g^{-1}$, as done in \cite{shankarread} for the $1+1$ dimensional $O(3)$ sigma model. In the large $g$ limit the local kinetic energy of the triad at each site dominates the intersite couplings in the Hamiltonian; i.e. this is a tight binding limit. It can be shown, for instance by computing in local coordinates on the configuration space of the triad \cite{avmthesis} the $Z_2$ character  $[\Lambda \{ {\bf e}_{A,B,C}({\bf x}_{m,n}) \}]^{2S}$ at each site that it is the action due to a monopole of strength $2S (mod)2$ at the origin of the solid sphere with opposite points identified (which is the configuration space of the local triad). In contrast to the case of a vector order parameter associated with antiferromagnets the monopole charge is $Z_2$ valued. As in the case of $SU(2)$ spins the effect of the monopole is to determine the quantization of the symmetric top formed by the dynamics of the triad ${\bf e}_{A,B,C}$ to be integer or half odd integer, according as $2S$ is even or odd. The eigenstates of a symmetric rotor are $(2J+1)^2$ fold degenerate; including the half odd integer sector they form the vector and spinor representations of $SO(4)$. Thus when $2S$ is odd the eigenstates of the free rotor are classified by left and  right angular momenta $(J_L,J_R)$ with both $J_L$, $J_R$ being equal integers. The ground state is a nondegenerate singlet formed by the mixing with neighbouring sites by the hopping term in the tight binding model, and the lowest excited states carry quantum numbers $(J_L,J_R)=(1,1)$ and are ninefold degenerate.  

When $2S$ is odd the spectrum of the half odd integer rotor at each site has a fourfold degenerate ground state with $(J_L,J_R)=(\frac{1}{2},\frac{1}{2})$; restricting the Hilbert space at each site to these four states (a valid approximation for large $g$) and coupling adjacent sites through the hopping term leads to an effective Hamiltonian
\begin{eqnarray}
H_e&=&-\lambda \sum_{<mn,m^{'} n^{'}>} \big{(} [\frac{1}{4} Tr M^{T}_{mn} M_{m^{'} n^{'}}]^2+\\ \nonumber
&&\frac{1}{2} Tr M^{T}_{mn} M_{m^{'} n^{'}} \big{)}
\end{eqnarray}
 where $M$ is a generator of $SO(4)$. Defining spin $\frac{1}{2}$ operators ${\bf S}_A$, ${\bf S}_B$ by $M^{0i}=\frac{1}{2}(S^i_A+S^i_B)$, $M^{jk}=\frac{1}{2} \epsilon ^{ijk} ((S^i_A-S^i_B)$ $H_e=- \lambda \sum_{<ij>} ({\bf S}_{Ai} \cdot {\bf S}_{Aj})( {\bf S}_{Bi} \cdot {\bf S}_{Bj})$. The four spin interaction term in $H_e$ favours the formation of a singlet between neighbours, and the nearest neighbour dimer limit of this model reduces to that studied in \cite{sondhi, moessner} and is disordered (also in agreement with the large$-N$ limit of \cite{sachdevlrgn}); the biquadratic coupling in $H_e$ in fact increases the tendency to disorder, as indicated for instance by numerical studies of spin orbital models on the triangular lattice \cite{spinorbital}, which show no tendency to order, unlike the Heisenberg model.  While the effective Hamiltonian $H_e$ differs from those studied in \cite{sachdevlrgn} by the presence of the biquadratic term it is likely that the two models are adiabatically connected at low energy.

  An important difference between phases obtained by disordering antiferromagnets and ordered states with noncollinear order is in the symmetry of the order parameter, which leads to distinction between the type of topological defects which can be formed. In \cite{haldane1} it was shown that when antiferromagnetic states are disordered by tunnelling events which change the Pontryagin index of the spin system from one time slice to the next they are associated with Berry phases which determine the nature of the disordered state depending on the value of $2S$ modulo the coordination number of the lattice. Since the second homotopy group $\pi_2 ( SO(3))$ vanishes (i.e., a skyrmion in a unit vector triad can be ccontinuously unwound to the trivial configuration) such effects are absent when noncollinear magnets are disordered. The presence of $Z_2$ vortices depends only on the fact that the order parameter is a rotation matrix; thus we expect their existence in more general systems, such as those disordered by fermion hopping, although the effects of the vortices will depend on the details of the model.  

In conclusion, a semiclassical approach to the description of the spin liquid phase of the model with ring exchange on the triangular lattice was presented. $Z_2$ vortices were shown to be generated, carrying a phase for the local order parameter which leads to a difference in the low energy physics for half odd integer and half even integer spin systems, and in the topological degeneracy of the ground state on surfaces with handles. 

 I thank F. D. M. Haldane and A. Ludwig for helpful discussions.

\end{document}